\documentclass{article}
\pdfoutput=1
\usepackage{mathtools}
\usepackage{amsmath}
\usepackage{amsfonts}
\usepackage{amsthm}
\usepackage{amssymb}
\usepackage{accents}
\usepackage{relsize}
\usepackage{upgreek}
\usepackage{hyperref}
\usepackage{enumitem}
\usepackage{subcaption}
\usepackage[utf8]{inputenc}

\usepackage{pictexwd,dcpic}
\usepackage{float}
\usepackage{graphicx}
\graphicspath{{./Image/}}

\usepackage{biblatex}
\bibliography{Discrepancy}

\theoremstyle{definition}
\newtheorem{definition}{Definition}

\DeclareMathOperator*{\argmin}{arg\,min}
\DeclareMathOperator*{\argmax}{arg\,max}

\DeclareMathOperator*{\E}{\mathbb{E}}
\DeclareMathOperator*{\cov}{Cov}

\renewcommand{\P}{\mathbb{P}}

\renewcommand{\epsilon}{\varepsilon}
\newcommand{\Reals}{\mathbb{R}}

\newcommand{\Uni}{\mathcal{U}^d}

\newcommand{\Nat}{\mathbb{N}}

\newcommand{\Normal}{\mathcal{N}}

\newcommand{\PM}[1]{\mathcal{P}\left( #1 \right) }
\newcommand{\PML}[1]{\mathcal{P}_{\ll \lambda}\left( #1 \right) }

\newcommand{\cp}{\theta}

\newcommand{\F}{\mathcal{F}}
\newcommand{\iid}{\underset{\mathrm{i.i.d}}{\sim}}
\renewcommand{\O}{\Omega}
\newcommand{\Time}{\mathcal{T}}
\newcommand{\Int}{\mathbb{Z}}

\newcommand{\kernel}{\mathcal{K}}

\newcommand{\von}{\mathcal{V}}
\newcommand{\dcc}[1]{\widehat{\Delta}_{#1}^T}

\newcommand{\scc}[1]{\widehat{\Delta}_{#1}^{T,\tau}}
\newcommand{\sccn}[1]{\widehat{\Delta}_{#1}^{T_n,\tau_n}}
\newcommand{\iscc}[1]{\Delta_{#1}^{T,\tau}}
\newcommand{\msd}{\overline{\Delta}_n}
\author{Pronko Nikita Konstantinovich}
\title{Change Point Detection with Optimal Transport and Geometric Discrepancy}
\begin{document}
\maketitle
\begin{abstract}
We present novel retrospective change point detection approach based on optimal transport and geometric discrepancy. 
The method does not require any parametric assumptions about distributions separated  by change points.
It can be used both for single and multiple change point detection and estimation, while the number of change points is 
either known or unknown.
This result is achieved by construction of a certain sliding window statistic  from which change points can be derived with elementary convex geometry in 
a specific Hilbert space.
The work is illustrated with computational examples, both artificially constructed and based on actual data.
\end{abstract}

\paragraph{Introduction}
Change point problem was firstly mentioned by Stewhart \cite{shewhart} as a problem of industrial quality control in the year 1928. 
This problem was solved by Girshick and Rubin \cite{girshick} in 1953. 
The optimal solutions for parametric  online formulation of the problem were provided by Shiryaev and Pollak \cite{shiryaev},\cite{pollak}.
Asymptotically optimal posteriori change point detection method was developed by Borovkov \cite{borovkov}. 
This method requires knowledge of parametric likelihood functions, however it  can be translated to nonparametric case with empirical  likelihood \cite{empirlike}.
The fundamental results in nonparametric study of change-point problem are due Brodsky and Darkhovsky \cite{brodsky}, and Horvath et al. \cite{horvath}. 
These methods are generally tailored  for one-dimensional data. 
Nonparametric change point inference with multidimensional data is  still an open problem. 
New methods are still being developed usually with applications of ideas from other scopes to change point data. 
For example, \cite{matteson}, where divergence from cluster network analysis is used in order to estimate
both the number and the locations of change points or \cite{lung}, where multivariate version of Wilcoxon rank statistic is computed.

This work is focused on nonparametric change point detection in multivariate data.
This problem has large scope of applications including  finance \cite{finance},genomics \cite{genomiccp} and signal processing \cite{signal}.

The problem of optimal transportation were introduced by Monge \cite{monge1781} in the year 1781. The modern statement of the problem is due to Kantorovich as it was stated in the seminal paper in 1941 \cite{kantorovich}.
We recommend \cite{villani2} and \cite{optimal} as modern references to the subject 

Vector rank and quantile functions were introduced by Serfiling \cite{serfling} as part Depth-Outlyingness-Quantile-Rank (DOQR) paradigm. Monge-kantorovich vector ranks were introduced by Chernozhukov et al. \cite{chernozhukov}.

Study of discrepancy were started by H. Weyl \cite{weyl} in 1916  with number theory as intended area of application, 
more computational aspects of discrepancy  were pioneered by Niederreiter \cite{nieder} (quasi- and pseudo- random number generation) and Hlawka \cite{hlawka} (numeric integration). 
We say that discrepancy is geometric when we talk about continuous distributions with fixed well-identified support, this is the opposition to the case of combinatorial discrepancy which study nonuniformity of finite distributions.
The main contribution to the theory of quadratic generalized geometric discrepancy is by Hickernel \cite{hicker}.
This is a work of a highly synthetic nature. 
This Synthetic nature is coming from the combination  of vector ranks and geometric discrepancy, 
optimal transport  and change point problems. 
All this concepts are rarely to never seen together in scientific publications, 
however in this paper they are intertwined in a singular computational methodology.

But why these results come to existence only now? 
The answer to this questions in our case is rooted into the fact that computing discrepancy used to be computationally challenging task for non-uniform probability distributions. 
However, in the recent publication \cite{chernozhukov} by Chernozhukov,Carlie, Galichon and Halin the Monge-Kantorovich vector rank functions were introduced. 
Vector rank function can be understood in the context of this work as homeomorphic extension of classical one-dimensional cumulative distribution functions to multidimensional spaces. 
The idea of using low-discrepancy sequences as a tool for the estimation of the vector rank functions was already present in the \cite{chernozhukov}. 
As the name suggests discrepancy is the natural measure of consistency of the sequence to be low-discrepancy sequence in the same way as the order statistic is the natural measure of the sequence to be ordered. 
Thus, as  classical one-dimensional ranks and order statistic is used in the construction of the classical Kolmogorov-Smirnov and Cr\'amer-Von Mizes tests of goodness-of-fit, 
the vector rank function and discrepancy can be used in order to construct natural extension of this tests in the multidimensional context.

Thus, the main objective of the work is the exploration of vector rank discrepancy as  methodology for nonparametric high-dimensional statistic.
We select change point problem as the test subject for this inquiry.

The strict measure theoretic approach mentioned above is strengthened in this work as we speak about probability measures 
instead of probability distributions and densities whenever is possible which contradicts current trends in the statistical science. 
This can be justified by convenience of understanding  measures as points in certain multidimensional spaces bearing intrinsic geometric structure. 
On the other hand, we view all mathematical constructions in this work as theoretical prototypes of certain concrete algorithms ad data structures  
which can be  implemented in order to solve practical tasks. 
Our approach to the problem is highly inspired by \cite{lindsay} and we use convex geometric intuition whenever possible.

\subparagraph{acknowledgement}
 Underlying research was conducted during author's work for master's degree in  "Mathematical methods of Optimization and Stochastics" program in NRU Higher School of Economics, Moscow, Russia.
 Author wants to thank head of the program Vladimir Spokoiny, his scientific advisor Andrei Sobolevski for suggesting the research topic and support, Alexey Naumov and Maxim Panov for the discussion in IITP RAS;
 Yuri Yanovich,Alexandra Suvorikova , Elena Chernousova and Alexey Kroshnin for fruitful discussion at the IUM seminar.   
 
\paragraph{Notation}

This is notation for the use in the sequel.

$\Reals^d$ is a $d$-dimensional euclidean space with $d < \infty$ .
$I = [0,1]$ is a unit interval and $I^d$ is a unit hypercube in $\Reals^d$. 
In line with Kolmogorov's axiomatics triple $(\O,\F,\P)$ is a probability space, and every observation as a random variable is assumed to be  a Borel measurable map from $\O$ to $\Reals^d$. For brevity $\Omega$ denotes the whole probability measure structure $(\O, \F, \P)$.

Every random variable $\xi : \O \to \Reals^d $ defines the probability measure $P = \xi_{\#} \P$ that is  referred to as the probability law of $\xi$, where $\xi_{\#} \P$ denotes a pushforward of the measure $ \P$ by $\xi$ . That is, for every Borel set $ B \subset \Reals^d$ it holds 
that $P(B) = \P(\xi^{-1}(B))$. In this case the notation $\xi \sim P$ is used. In case $\xi$ is a random process over domain $\Time$ notation $\xi \sim P$ 
means that $P$ defines the hierarchical  probability distribution system of $\xi$ as a whole. 
In this case for all $t,s \in \Time$ the measure $P_t$ is the distribution of a single observation $\xi_t$ and $P_{t,s}$ 
is the joint distribution of $\xi_t$ and $\xi_s$ i.e. $\xi_t \sim P_t, (\xi_t, \xi_s) \sim P_{t,s}$. 
$ \xi \iid P  $ means that $\xi$ is independently identically distributed sample  (i.i.d) with probability law $P$.  

A set of all probability laws with finite first and second moments is denoted by $ \PM{\Reals^d} $.
A subset of $\PM{\Reals^d}$  of probability measures absolutely continuous with respect to Lebesgue measure $\lambda$ is denoted by $\PML(\Reals)$. $\Uni$ is the uniform distribution over $I^d$. It is obvious that $\Uni \in \PML{\Reals^d}$.In the sequel $I^n$ is assumed to be implicitly equipped with $\mathcal{U}^n$ for each $n \in \Nat$, which makes it into a probability space,  and  $I^{0} = \{ \emptyset \} $ is equipped with the counting measure.

If $\xi_n$ is an  sequence of random elements in some metric space with distance metric $\rho$, when it is said that $\xi$ converges to $x$ in probability if for each $\epsilon \in \Reals_{++}$ it
holds $ \lim_{n \to \infty} \P( \rho(\xi_n, x) >  \epsilon) = 0 $, which is denoted as  $\xi_n \to_{\P} x$. In case  $\rho(\xi_n,x)$ is not a measurable function for all $n$ the convergence in outer probability may be introduced and denoted by $\xi \to_{\P \bullet} x $. Having $A^\epsilon_n = \{ \omega \in \Omega : \rho(\xi_n(\omega),x) > \epsilon \}$ the convergence in outer probability is equivalent to $\lim_{n \to \infty} \inf \{ \P(B) | B \in \F : A_n^\epsilon \subset B \} = 0 $ for all $\epsilon > 0$.   

If $\xi$ is a discrete time ergodic process with convergence in probability, 
\begin{equation}
\label{Ergodic}
 \frac{1}{n} \sum^n_{i = 1} f(\xi_i) \to_\P \int_{\Reals^d} f \, \mathrm{d} \bar P
\end{equation}
for every bounded Lipschitz function $f$, we say that $\bar P$ is the weak limit distribution of 
$\xi$.
\paragraph{Problem Formulation}
In this section a brief review for common variations of change point problem is provided.

Ordered Set $\Time$ is associated with time. It is possible to discuss change point problem with continuous time \cite{dvoretzky}. However, this work deals only with case of discrete time. It is also possible to consider a change point detection for random fields  \cite{brodsky}.

Two main types of change point problem are offline and online detection. In offline detection  $\Time$  is assumed to be a finite set of size $|\Time| = T$. 
Without loss of generality, it is assumed that $\Time = \{1, \ldots, T \}$. Initial observations $X$ are treated as a time series indexed by $\Time$. 
Firstly,  we discuss the case of a single change point.  
\begin{definition}
The \textbf{change point} of $X$ is an unknown moment of time $\cp \in \Time$ such that $ (X_{t})_{ t = 1 }^\cp  \sim P^1$ 
 and $(X_{t })^T_{t = \cp + 1}  \sim P^2 $. There are two possibilities. Firstly, both subsamples may be i.i.d with  $(X_t)^\cp_{t = 1} \iid \bar P^1 $ and
  $(X_t)^{T}_{t = \cp + 1} \iid \bar P^2$.  It is also possible to consider unknowns $\bar P^1$ and 
  $\bar P^2 $ to be ergodic weak limit distributions of  $(X_t)^\cp_{t = 1} $ and
  $(X_t)^{T}_{t = \cp + 1}$  respectively.
\end{definition}  
   Then offline single change point detection problem is a hypothesis test for $H_0 : \bar P_1 = \bar P_2$ versus $H_1 : \bar P_1 \neq \bar P_2$ 
   for every possible estimate $\hat \cp $ of $\cp$ having $1 < \hat \cp < T$.  Retrieving valid value of $\hat \cp$ is a related problem which will be referred as a change point estimation.
 \begin{definition} 
  In case of \textbf{multiple change points}  existence of up to $N$ change moments 
  $ \cp_1  < \ldots < \cp_n  < \ldots \cp_N \in \Time $ is assumed. In simpler formulation of problem value of $N$ is assumed to be known, while in more complicated one $N$ is also an object of estimation. 
	 As the previous variation  the problem is a hypothesis test about distribution of samples 
  $(X_t)_{t = \cp_{n - 1} + 1}^{\cp_{n}} \sim \bar P_n $ where $\cp_0 = 1$ and $\cp_{N + 1} = T$ for simplicity. 
	 The problem splits into multiple hypothesis tests for $H^n_0 : \bar P_n = \bar P_{n + 1}$ versus $H^n_1 : \bar P_n \neq \bar P_{n + 1}$ for all possible estimates  $\hat \cp$ of $\cp$ in case of known $N$. 
	 For unknown $N$ the problem is structured around testing against $H_0 :\bar P_1 \neq \bar P_2  \neq \ldots \neq \bar P_{\hat N}$ for all estimates 
	 $( \hat N, \hat \cp)$ of $(N,\cp)$ where possible values of $\hat N$ are  constricted to some meaningful finite set.
\end{definition}
  
For online change point problem  $\Time \cong \Int$ as an ordered set, and $X$ can be thought as infinite times series unfolding in  real time.
In this case the goal is to identify change point  $\cp$ as soon as possible which means using minimal amount of observations $X_t$ with $t \ge \cp$.
Branding every moment of time as a change point will achieve zero delay. However, this detection procedure is unacceptable as it achieves maximal false alarm rate. 
This means that change-point detection algorithm needs to be tuned in for minimal delay with fixed false alarm rate.

  In a more theoretical framework  a single change point $\cp_0$ can be considered, which implies that infinite set $\{ X_t : t < 0 \}$ of observations with prior distribution can be used for detection. 
  However, in more practical situation the stream of chanege ponts $\cp : \Nat \to \Time$ is considered, with only finite sample $\{ X_t : \hat \cp_{n - 1} < t \le \cp_n \}$ available for detection of $\cp_n$.

Generally, methods of change point detection are focused on detection of change in such properties of statistical distributions as mean, variance and spectral density.
The method presented in this work is focused on change in ergodic  weak limit distribution of data. 
Which means that change in mean and variance must be detectable, although, 
change in spectral density is not. 
Moreover, changes restricted to such properties of distribution 
as median or higher moments also must be traceable.
Another restriction of proposed solution is a demand for all measures 
$\bar P_n$ to belong to $\PML{\Reals^d}$.

Construction of proposed change point detection starts with  offline single detection problem.
 We restrict our attention to models with i.i.d distributed observations along one side of change point.
 In order to give meaningful characteristic to point estimates of change point this problem can be reformulated as certain probabilistic game against Nature. 
 Two  sample spaces $(\O^1, \F^1, \mathbb{P}^1)$ and $(\O^2, \F^2, \mathbb{P}^2)$ with corresponding families of random variables $\xi^1$ and $\xi^2$ taking values in $\Reals^d$ are assumed to exist. We assume that both families
  are intrinsically i.i.d which means that $\xi^1 \iid P_1$ and $\xi^2 \iid P_2$ and the same holds for $\xi^2$. This means that both sample probability spaces can be decomposed as $\Omega^1 = \prod^\infty_{t = 1} \Omega^1_t$ and 
  $
  \Omega^2 = \prod^\infty_{t = 1} \Omega^2_t
  $  
  such that for each $\omega \in \O^1$  the random variable $\xi_{t}^1(\omega)$  depends only on 
  $\omega^t$.
 The common sample space is constructed by the rule
\begin{equation}
\Omega_t = \Omega^1_t \sqcup \Omega^2_t 
= \big(  \{1\} \times \Omega^1_t \big) \cup  \big( \{ 2 \} \times \Omega^2_t \big),
\quad
\Omega = \prod^\infty_{t = 1} \Omega_t 
\end{equation} 
with $\sigma$-algebras generated by a set  $ \{ \{1 \} \times A : A \in \F_{1}  \} \cup  
\{ \{2 \} \times B : B \in \F_{2}  \}$ . Note, that the only possible probability distribution on  
the set $\{1,2 \}$ is  a Bernoulli   distribution $\mathrm{Bern}(r)$ with a parameter $r \in [0,1]$. So probability measure on $\Omega$ can be defined by $ \mathbb{P}(S) = r\mathbb{P}_1(\iota^{-1}_1 S) + (1 -r)\mathbb{P}_2(\iota^{-1}_2 S)$, where $\iota_j$ is a natural injection defined by $ \omega \mapsto (j,\omega)$.  Nature defines value of $r$ and distribution laws  $P_1$ and $P_2$. Random variables $(b_t,X_t)$ are constructed by 
letting $(b_t,X_t)(j, \omega) = ( j, \xi^j_t(\omega) )$. Then, nature generates sample of $T$ observations of $(b,X)$, sorts it by $b$, then erases values of $b$ which retains observable sample of $X$. Observer either claims that there were no change point which leads to a victory in case 
$ P_1 = P_2 $ or equivalently $r$ is too close to either $0$ or $1$. Otherwise, the observer gives point-estimate $ \hat r$ of $r$ which can be converted to a classical change-point estimate $\hat \cp = \hat r T $. 

This construction also can be understood as a Bayesian assumption $\cp \sim \mathrm{B}(T,r)$, where $\mathrm{B}(T,r)$ stands for binomial distribution. 
However, the change point problem is a single element sample problem, which means that only minimal properties of the parameter can be inferred from the data. This does not change the problem except for treating $\hat \cp$ as an estimate of the mean. 

Obviously,the  observer chooses strategy which maximizes chances of victory. This can be interpreted in terms of loses $L$, which are equal to $L(0) = \min \{  1 - r, r \}  = L(1)$ in case observer claims no change point and $ L(\hat r) =  | \hat r - r  | $ otherwise.
Let  $ L(\hat r) = \min{1 - \hat{r}, \hat{r}}$  in case the observer gives estimate when no real change point exists.
 
 If one preserves $r$ while increasing $T$ the empirical distribution of $X$ will converge to a measure $\mu_r = r \bar P_1 + (1 - r) \bar P_2$, the convex combination of measures. It possible to think of $\mu_r$ as contained inside one-dimensional interval $ [\bar P_1, \bar P_2]$ embedded into the infinite-dimensional convex space of probability measures. In order to construct such measures from random variables we will use random mixing map $I^s_Z : Y \mapsto b_sY + (1 - b_s)Z$ in the sequel. Here $b_s \sim \mathrm{Bern}(s)$ and $s \in [0,1]$.

In order to propose the solution strategy for  the change point problem we  introduce vector rank function  $ R : \Reals^d \to \Reals^d$ such that $R_{\#} \mu_r = \Uni$, a known reference distribution of simple structure. For our application $\Uni$ is a uniform distribution over a unit hypercube $I^d$. It was shown in \cite{chernozhukov} and \cite{decurninge} that $\hat R$, an estimate of $R$ , can be recovered from data without any parametric assumptions. 
 
If data is i.i.d distributed with laws $P_1$ and $P_2$ on each side of change point, when  figure 1 will  exhibit relations between probability distribution laws discussed so far. 
\begin{figure}[H]
$$
\begindc{\commdiag}[30]
\obj(0, 40)[PP1]{$\P^1$}
\obj( 40,40)[PP2]{$\P^2$}
\obj( 20,40)[PP]{$\P$}
\obj(0, 20)[P1]{$P_1$}
\obj(20,20)[M]{$\mu_r$}
\obj(40,20)[P2]{$P_2$}
\obj(0, 0)[RP1]{$R_\# P_1$}
\obj(40, 0)[RP2]{$R_\# P_2$}
\obj(20, 0)[U]{$\Uni$}
\mor{P1}{M}{$I^r_{\xi^2}$}
\mor{P2}{M}{$I^{1-r}_{\xi^1}$}[\atright,\solidarrow]
\mor{PP1}{PP}{$I^r_{\bullet} $}
\mor{PP2}{PP}{$I^{1-r}_{\bullet}$}[\atright,\solidarrow]
\mor{P1}{RP1}{$R$}
\mor{P2}{RP2}{$R$}
\mor{RP1}{U}{$I^r_{R (\xi^2)}$}
\mor{RP2}{U}{$I^{1 -r}_{R (\xi^1) }$}[\atright,\solidarrow]
\mor{M}{U}{$R$}
\mor{PP1}{P1}{$\xi^1$}
\mor{PP2}{P2}{$\xi^2$}
\mor{PP}{M}{$X$}
\enddc
$$
 \caption{  Diagram with measures as objects and pushforwards as arrows . The top row represents abstract measure spaces which are pushed forward to unknown measures by observed Random variables. The bottom row exhibits state of the probability laws 
after application of the vector rank function of $\mu_r$.} 
\end{figure}
The diagram shows that application of vector rank function can be thought as a probabilist's "change of basis". The pragmatic value for change point problem is in elimination of unknown distribution $\mu_r$ from the model.  

It is obvious that the diagram depicted at fig. 1 commute. As the pushforward acts as a linear map of signed measures defined over two different measurable spaces. If $\alpha$ and $\beta$ are measures, $x, y \in \Reals$, $B$ is a measurable set and $f$ is a measurable function, then  
 $$
  f_{\#}(x\alpha + y\beta)(B) = x \alpha( f^{-1}(B))  + y \beta(f^{-1}(B))
  = x f_{\#}\alpha (B)    + y f_{\#}\beta(B).
 $$  
 Therefore, by linearity of the pushforward
 $$
 \Uni = R_\# \mu_r =   R_\# ( r P_1 + (1 -r) P_2) = r R_\# P_1 + (1 - r)R_\# P_2 . 
 $$
 Thus, the diagram is correct.

The real result of change-point estimation will depend on measurable difference between $P_1$ and $P_2$. We introduce notion of Kullback-Leibler divergence in order to quantify this difference.
\begin{definition}
\textbf{ Kullback-Leibler divergence} between measures $P_1$ and $P_2$ admitting densities $p_1$ and
and  $p_2$ respectively is
$$
 \mathrm{KL}(P_1 , P_2) = \int_{\O_1} \log \frac{p_1}{p_2} \, \mathrm{d}P_1 
$$
\end{definition}
Note  that while $P_1,P_2  \in \PML{\Reals^d}$ they admit densities. Now we formulate a sufficient condition on $R$ that ensures that our transformations preserve divergence between distributions.

If $R$ is invertible almost everywhere $P_1$, when $\mathrm{KL}(R_{\#}P_1,R_{\#}P_2) 
 = \mathrm{KL}( P_1,P_2) 
$
 Let $X$ be a random variable such that  $X \sim P_1$. Then, $R(X) \sim R_{\#} P_1$ and by
  invertibility  a.e. of $R$  distribution $R_\# P_1$ has density $p_1 \circ R^{-1}$ a.e. ; similar fact is also true for $P_2$ .
  Then,
 $$
  \mathrm{KL}( R_\# P_1, R_\# P_2   ) = \E \log \frac{ p_1(  R^{-1}  R  X) }{ p_2 ( R^{-1}  R  X )  }  = \E \log \frac{ p_1 (X) }{ p_2  ( X)  } =  \mathrm{KL}( P_1,P_2) 
 $$
\paragraph{Formulation of the method}
In this chapter our approach to construction of the vector rank function's estimates $\hat R_n$ is presented. The main goal of This approach is  the avoidance of approximation of the entire function $R$. 
Note, that even the estimation that the aim of this task is actually not a precise estimation of values $R(X_i)$ for each observation $X_i$ but a construction of 
an estimate that preserves the relative geometry of a sample $X$ along its change points.

The work \cite{chernozhukov}  presents continuous, semidiscrete, 
and discrete methods of estimation of $R$. 
The Implementation of change point detection discussed in this section is based solely on discrete approach. The reason behind this choice is computational simplicity.

Vector rank function in general can be understood as $(P,\,mathbb{U})$-almost surely homeomorphisms  between supports of sample probability distribution $P$ and reference probability law of choice $\mathbb{U}$
($R$ is defined almost everywhere $P$ and continuous with respect to subset topology, and admits similar continuous inverse existing almost everywhere $\mathbb{U}$)
, 
such that  $R_\# P = \mathbb{U} $ and both depth median and symmetries are preserved. 
This symmetries  and the concept of depth median need to be additionally defined, 
which goes beyond of the scope of this paper. In our application uniform measure over unit hypercube $I^d$ is selected as the reference $U$.
In case $d = 1$ the cumulutive distribution function (cdf) of $P$ is also the 'vector' rank function of the probability distribution of $P$.

The vector rank function is not unique in general. We use Monge-Kantarovich vector rank developed in \cite{chernozhukov}, which is defined as optimal transport between 
$P$ and $U$
\begin{equation*}
R = \argmin_{R : R_{\#}P = \Uni } \int_{\Reals^d} d^2(R(x),x) \, \mathrm{d}P(x).
\end{equation*}
Vector rank role of the optimal transport map $R$ can be intuitively justified by the equivalence of the above optimization problem to the maximization of the integral
\begin{equation*}
	\int_{\Reals^d} \langle R(x) - m(\Uni) , x - m(P)  \rangle \, \mathrm{d}P(x),
\end{equation*}
where $m(\Uni)$ and $m(P)$ stands for depth medians of distributions $\Uni$ and $P$ respectively. 
Thus, the optimal transport preserves geometric properties of distribution $P$ which can 
be expressed by inner products of points relative to its center of symmetry. This is what is understood as the relative geometry of the data 
,and what we try to preserve during  vector rank estimation. 

In this work we are focused on the discrete estimation of vector rank function.
Let $u$ be an equidistributed sequence of points in $I^d$. That is, for every Lipschitz continuous function  $f$ defined on $I^d$ it holds
\begin{equation}
\label{eqidist}
 \lim_{n \to \infty} \frac{1}{n} \sum^n_{i = 1} f(u_i) = \int_{I^d} f(x) \, \mathrm{d}x.  
\end{equation}
Then, for $T$ observations define $Y_i = \hat R_T(X_i) = u_{\sigma^*(i)}$, where 
\begin{equation}
\label{opt_assignment}
 \sigma^* = \argmin_{\sigma \in S^T} \sum^T_{i = 1} \|  X_i - u_{\sigma(i)}    \|_2^2.
\end{equation}
In case convergence in \eqref{eqidist} is understood as convergence a.s or even as convergence in probability the sequence $u$ can be taken as an i.i.d  
sequence of random variables  or as an ergodic process with a uniform weak limit distribution. 
However, this  implementation is more suited for  deterministic form of $u$. 
Discrepancy can be understood as a natural measure of slackness of condition \eqref{eqidist} for a fixed value of $T$.

Classical (one-sided) geometric discrepancy of the sequence $(u_i)^n_{i = 1}$ is described by the formula
\begin{equation}
	\label{Disc1}
	D(u) = \max_{x \in I^d} \left| \frac{\big| \{ u_i | 1 \le i \le n\} \cap [0,x] \big|}{T}  - \prod^d_{j = 1} x^j    \right|.
\end{equation}
Note that \eqref{Disc1} admits representation $ D(u) = \| \varphi_u \|_\infty $ for a certain function $\varphi_u$. 
This idea were used in \cite{hicker} to introduce generalized quadratic discrepancy with supremum-norm replaced by a 
certain quadratic norm in a certain Sobolev Space. The result can be expressed  as
\begin{multline}
	\label{Disc2}
	\Big(D_2^\kappa(u) \Big)^2 = 
	\\ =
	\iint \eta(x,y) \, \mathrm{d}x \mathrm{d}y - \frac{2}{n} \sum^n_{i = 1} \int \eta(x,u_i) \, \mathrm + \frac{1}{n^2}\sum^n_{i,j = 1} \eta(u_j,u_i),
\end{multline}
where $\eta$ is the reproducing Hilbert kernel of the Sobolev space 
\begin{equation}
	\eta(x,y) = \prod^d_{i = 1} \Big( M + \beta^2\big( \kappa(x^i) + \kappa(y^i) + \frac{1}{2}B_2\big( x^i - y^i \mod 1 \ \big) + B_1(x^i)B_1(y^i) \Big)
\end{equation}
with $\beta \in \Reals$ standing for  a scale parameter and $\kappa $ standing for a functional parameter with square-integrable derivative  with $\int^1_0 \kappa = 0$ and the value $M$ defined by
$$
M = 1 - \beta^2 \int^1_0 (\kappa')^2,
$$
and $B_i$ is the ith Bernoulli polynomial. 
Note, that in case  $d = 1$ statistic \eqref{Disc1} turns into Kolmogorov-Smirnov statistic and \eqref{Disc2} turns into Cr/`amer-Von Mizes statistic for uniform distribution test.

The function $\kappa$ is a functional parameter defining exact form of the discrepancy. 
In this text we use star discrepancy produced by
$$
\kappa^* = \frac{1}{6} - \frac{1}{x^2}
$$
and the centred discrepancy produced by selecting
$$
\kappa^c = -\frac{1}{2} B_2\left( x - \frac{1}{2}  \mod 1\right)
$$

The case of $u$ being a low-discrepancy sequence is a particularly well suited for our application.
\begin{definition}
\textbf{low-discrepancy sequence} is a deterministic sequence $u$ in $I^d$ designed with a goal of minimizing value the $D^\kappa_p(u_i)^n_{i = 1}$ for each natural number $n$.
\end{definition}
Definition above is rather informal. However, it is postulated firmly that for any low-discrepancy sequence $u$ property \eqref{eqidist} holds and $\lim_{n \to \infty} D^\kappa_p(u_i)^n_{i = 1} = 0$ for any choice of $p$ and $\kappa$. 
Thus,  the rate of convergence of $D^\kappa_p(u_i)^n_{i = 1}$ can be understood as a measure of efficiency of a low-discrepancy sequence $u_i$.

In our application we use Sobol sequence with grey code implementation. Sobol sequence were introduced by Sobol \cite{sobol} and the grey code implementation is due \cite{antonov}. 
Detailed investigation of the nature of this sequence goes beyond the scope of this work. However, any other low-discrepancy sequence can be used.

For a Sobol sequence value $D^\kappa_2(u_i)^n_{i = 1}$  converges to zero as $O(n^{-1 + \epsilon})$, 
where $\epsilon$ depends on $d$ implicitly. While convergence rate of discrepancy  for a random uniform sequence  is $O(1/\sqrt{n})$, for $d \ll 200$ value  of $ \epsilon < 1/2 $, 
which makes low-discrepancy sequence rather efficient.  
However, scrambling and effective dimension techniques can increase rate of convergence \cite{hickernell2}.

It can be established that the sequences $u$ and $X$ have no repeating values almost surely, 
As our data is assumed to come from atomless distributions. 
This means that the problem of finding Permutation $\sigma^*$ in \eqref{opt_assignment} is the optimal assignment problem. 
The Optimal assignment problem  is the special case the linear programming, 
which can be solved in $O(n^3)$ time by the Hungarian algorithm \cite{kuhn}\cite{tomizawa}. 
Amortizations based on applications of parallel programming can improve computation complexity to $O(n^2 \log n)$. 
Approximate algorithms can be used for faster computations, for example \cite{Kuturi}.

Note, that even if sample $X$ has an i.i.d distribution, then the  resulting transport $Y$ is not  independent itself. 
However, the covariance of the elements is converging to zero as $T$ goes to infinity. 
Let $n \neq m$ be two indices less or equal  to $T$ and $i,j$ be coordinate indices. As $(X_k)^T_{k = 1}$ is assumed to be i.i.d, it follows that $Y^i_n$ has a discrete uniform distribution over $\{ u^i_k \}^T_{k = 1}$. 
Then, 
\begin{multline*}
\cov(Y^i_n,Y^j_m) = \E Y^i_n Y^j_m - \E Y^i_n \E Y^j_m  = \sum^T_{k = 1} \sum^T_{ l \neq k} \frac{u^i_k u^j_l}{T(T - 1)} - \sum^T_{k = 1} \sum^T_{l = 1} \frac{ u_k^i u_l^j }{T^2} =
\\ =
\sum^T_{k = 1} \sum^T_{ l \neq k} \frac{u^i_k u^j_l}{T^2(T - 1)} - \sum^T_{k = 1} \frac{ u^i_k u^j_k}{T^2},
\end{multline*}
so by bounding from above and below with equidistribution property of $u$ in the limit case
\begin{equation*}
 \cov(Y^i_n,Y^j_m) \le  \sum^T_{k = 1} \sum^T_{l = 1} \frac{u^i_k u^i_l}{T^2(T - 1)}  \xrightarrow[T \to \infty]{} \lim_{T \to \infty} \frac{\E U \E U}{T - 1 } = 
\lim_{T \to \infty} \frac{ 1}{4(T - 1) } = 0, 
\end{equation*}
\begin{equation*}
 \cov(Y^i_n,Y^j_m) \ge  - \sum^T_{k = 1} \frac{ u^i_k u^j_k}{T^2} \ge    - \sum^T_{k = 1} \frac{ u^i_k }{T^2}  \xrightarrow[T \to \infty]{} \lim_{T \to \infty} - \frac{\E U }{T  } = 
\lim_{T \to \infty} - \frac{ 1}{2T } = 0;  
\end{equation*}
 where $U$ is a uniform random variable on $[0,1]$. Thus, the value $\cov(Y^i_n,Y^j_m)$ converges to zero. As variance of $Y^i_n$ converges to the variance of a standard uniform distribution on $[0,1]$  
 we will treat sample $Y$ as uncorrelated in asymptotic context under assumption of i.i.d. distribution of $X$.

 It can be easily seen that that quadratic discrepancy admits a degenerate  V-statistic representation  with the kernel $\kernel$:
 \begin{equation}
	 \Big( D^\kappa_2(y) \Big)^2 =  \frac{1}{n^2} \sum^n_{i,j = 1} \kernel(y_i,y_j), 
 \end{equation}
 assuming sample $y$ of $n$ elements. Properties of V-statistic produces the  asymptotic result 
 $$
       n \Big( D^\kappa_2(y) \Big)^2  \xrightarrow[n \to \infty]{d} V = \sum^\infty_{i = 1} \lambda_i Z_i^2 \sim \von(\lambda),
 $$
 where  $Z \iid \Normal(0,1)$ and $\lambda$ are non-zero eigenvalues of the integral operator $\mathcal{A}$ defined by the relation
 $$
  \mathcal{A}(f)(x) = \int^1_0 f(y)\kernel(x,y) \mathrm{d}y 
 $$
 It can be shown that $\mathcal{A}$ is in fact positive-semidefinite  and  trace-class, which means that all $\lambda_i > 0$ and that
 $$
 \sum^\infty_{n = 1} \lambda_i < \infty
 $$
 Eigenvalues of $\mathcal{A}$ can be approximated by a  finite collection of $N$ numbers $\hat \lambda$ with Nystr\"om method.
 As it was shown in \cite{choirat} the twofold approximation of cdf of $V_{|N} = \sum^N_{i = 1} \widehat{\lambda}_i^N Z_i^2$ is possible for fixed natural numbers $N$, $K$ and parameter $\alpha \in (0,1]$:
\begin{multline}
\label{numeric}
 \P(V_{|N} < x) \approx \\ \approx \frac{1}{2} - \sum^K_{k = 0} \frac{\sin\left(
 \frac{1}{2}\sum^N_{1 = 1} \arctan\left(2\left(k + \dfrac{1}{2} \right) \alpha \widehat{\lambda}^N_i\right) - \left(k +  \frac{1}{2} \right) \alpha x  \right) }{\uppi \left( k + \frac{1}{2} \right) \prod^N_{i = 1}\sqrt[4]{1 +  \left(2\left(k + \frac{1}{2} \right) \alpha \widehat{\lambda}^N_i \right)^2 }}.
\end{multline}
This formula can be used for computing quantiles  and critical values of $\von(\lambda)$ with simple one-dimensional zero-seeking algorithm.
\subparagraph{Case of A Single Change Point }
In this chapter an approach for detecting a single change point is presented. We impose a model assumptions that for a fixed $\tau$ change point $\cp$ either satisfy $\tau <  \cp  < T - \tau$ or it does not exist .
This value $\tau$ can be selected in a such way that $1 \ll \tau \ll T$  and be used as a sliding window bandwidth defining  a 'control chart'-like object which we refer to as diphoragram.
\begin{definition}
 \textbf{empirical sliding diphoragram} for change point data $(X_t)^T_{i = 1}$ is defined by
$$ \widehat{\Delta}_t^{T,\tau} = \Big( D^\kappa_2\big(\hat R_T(X_i)\big)^{t + \tau}_{i = t} \Big)^2 
 =  \Big( D^\kappa_2(Y_i)^{t + \tau}_{i = t} \Big)^2 
$$
and \textbf{ideal  sliding diphoragram} by
$$
    \Delta_t^{T,\tau} = \Big( D^\kappa_2\big( R_{\mu_r}(X_i)\big)^{t + \tau}_{i = t} \Big)^2 
 $$
 for $t$ in range from $1$ to $T'_\tau = T - \tau$.
 \end{definition}
 If $\kappa$ is a continuous function, then  $ \scc{t} \xrightarrow[]{P} \iscc{t} $ as $T \to \infty$.
 With this condition discrepancy is a continuous function of data. So, convergence of vector ranks proved in \cite{chernozhukov} implies convergence of diphoragrams.
 Note, that with kernel representation  charts admit a difference representation
$$
 \scc{t + 1} -\scc{t} =  \frac{1}{\tau^2} \left(\sum^{t+1 + \tau}_{i = t + 1} \kernel(Y_i,Y_{t + \tau + 1})
 -  \sum^{t+ \tau }_{i = t } \kernel(Y_i,Y_{t}) \right).
$$
From the computational standpoint this means that computation of the whole time-series $\dcc{}$ and $\scc{}$ takes only quadratic time $O(dT^2)$ in number in observation. 
Moreover, if crisp optimal transport with low-discrepancy sequence $u$ is used in estimation of vector rank function, then $A_Y = A_u$, so all values can be precomputed. 

For  the application to the change point problem  consider two increasing sequences of integers  $T_n$ and $\tau_n$ such that $ \frac{T_n}{\tau_n} =  a > 1 $ for all $n \in \Nat$ are constructed.  
Then as $Y_i$ are assumed to be independent $\sccn{t}$ and $\sccn{t + \tau_n}$ are also independent random variables. 
\begin{definition}
\textbf{mean sliding discrepancy} for set sample $(Y_i)_{i = 1}^{T_n}$ is computed by
$$
 \msd = \frac{\tau_n^2}{T_n}\sum^{a - 1}_{j = 0} \sccn{1 + j\tau_n}
$$
\end{definition}
Note, that the mean sliding discrepancy is not the same  as the discrepancy of the whole sample. By independence, in case $H_0$ holds, as $n \to \infty$ 
$$
\msd \xrightarrow[T \to \infty]{d} \sum^a_{j = 1} \sum^\infty_{i = 1} \frac{\lambda_i}{a} Z_{i,j}^2 \sim \von\left( \left( \frac{\lambda_i}{a} \right)^a_{j = 1} \right)^\infty_{i = 1},
$$
where $Z \iid \Normal(0,1)$. Otherwise, there will be a sliding discrepancies $\sccn{t}$ sampled form the non-uniform data which goes to $\infty$ in probability as $n \to \infty$. 
So, the whole sum $\msd \xrightarrow[n \to infty]{\P} \infty$. 

In case  the single change point exists the statistic $\sccn{}$ is expected to attain the minimal value at such moment of time $t^*$ when the  empirical distribution  of $(Y_t)^{t^* + \tau_n}_{t = t^*}$ 
approaches the empirical distribution of the whole sample, as the empirical distribution of the whole sample is converging to the $\Uni$. Let 
       $$t^*_n = \argmin_{1 \le t \le T_n - \tau_n } \sccn{t}.$$
It is expected that  the ratio of numbers of elements from both sides of change point  in the subsample used in the computation of $\sccn{t^*_n}$ 
and in the whole sample are equal. This can be represented in the algebraic relation
$$
 \frac{ \tilde{\cp}_n - t^*_n  }{\tau_n}  = \frac{\tilde{\cp}_n}{T_n} = \hat r_n,
$$
where $\tilde \theta_n$ is the produced estimate of the change point. This provides the expression for the estimate
$$
 \tilde \theta_n =  \frac{t^*_n}{1 - a^{-1}}.
$$
We accept $H_0$ for a fixed $\gamma$-level in case  
\begin{equation} 
\label{pval}
p_n = \P\Big( V_{|N} \le \msd \Big) <   1 - \gamma,  
\end{equation}  
where the cdf is numerically estimated as in \eqref{numeric} for some fixed parameters $N$ and $K$. Otherwise reject $H_0$ and state that there was a change most probable at $\hat \cp_n$. 

In order to reason about properties of the estimate $\hat \cp_n$. Let $\mathcal{M}(I^d)$ denote  space of finite signed measures other the hypercube $I^d$.
\begin{definition}
\textbf{Space of nonuniformities} $\mathcal{D}$ is  defined as a quotient of the real vector spaces
\begin{equation*}
 \mathcal{D} = \frac{\mathcal{M}(I^d)}{\Reals \Uni}
\end{equation*}
endowed with a Hilbert space structure by inner product defined for $ [\nu], [\mu] \in \mathcal{D}$ by 
\begin{equation}
\label{nonun_inner_product}
\big\langle [\nu], [\mu] \big\rangle =  \int \int \kernel(x,y)  \, \mathrm{d} \nu(x) \, \mathrm{d} \mu(y)
\end{equation}
\end{definition}
Note, that the relation \eqref{nonun_inner_product} is well defined as for all $a,b \in \Reals$
\begin{multline*}
\Big\langle \big[\nu + a \Uni\big], \big[\mu + b\Uni\big] \Big \rangle  =
 \int \int \kernel(x,y)  \, \mathrm{d} \nu(x) \, \mathrm{d} \mu(y) +
 b\int \int \kernel(x,y)  \, \mathrm{d} \nu(x) \, \mathrm{d} y + \\ + 
 a \int \int \kernel(x,y)  \, \mathrm{d} x \, \mathrm{d} \mu(y) +
 ab \int \int \kernel(x,y)  \, \mathrm{d} x \, \mathrm{d} y 
 = \int \int \kernel(x,y)  \, \mathrm{d} \nu(x) \, \mathrm{d} \mu(y),
\end{multline*}
as $ \int \kernel(x,y) \, \mathrm{d} x = 0  $ for any value of $y$ and $\kernel$ is symmetric and is indeed an inner product as $\kernel$ is a positive-definite kernel. 
As the line $\Reals \Uni$ intersects simplex of probability measures $\mathcal{P}(I^d)$ only in one point ($\Uni$ itself) the natural projection $\mu \mapsto [\mu]$ is injective on $\mathcal{P}(I^d)$, 
so we denote a nonuniformity arising from each $\mu \in \mathcal{P}(I^d)$ just as $\mu$.   
If every subsample $(Y_i)_{i = t}^{t + \tau}$ is associated with an empirical measures $\hat \mu_t = \sum_{i = t}^{t + \tau_n } \delta_{Y_i}$, then $$\sccn{t} = \| \hat \mu_t  \|^2_{\mathcal{D}} = \langle \hat \mu_t,  \hat \mu_t \rangle.$$ 
By construction of the vector rank function $ \| r R_{\#} P_1 + (1 - r) R_{\#} P_2 \|_{\mathcal{D}} = 0$, hence $  r R_{\#} P_1 + (1 - r) R_{\#} P_2 = 0 $ in $\mathcal{D}$.  This produces result
\begin{multline}
\label{geometric_idea}
 R_{\#} P_2 = -\frac{r}{1 -r} R_{\#} P_1 \Longrightarrow  \Big\|   R_{\#} P_2  \Big\|_{\mathcal{D}}^2  = \frac{r^2}{(1 - r)^2} \Big\| R_{\#} P_1 \Big\|^2_{\mathcal{D}}   \Longrightarrow  \\
	\Longrightarrow
	r = \frac{\| R_\# P_2 \|_{\mathcal{D}}}{\| R_\# P_1 \|_{\mathcal{D}} + \| R_\# P_2 \|_{\mathcal{D}}}
\end{multline}
Considering that $r = 1/2$ it follows that $ [R_{\#} P_1]  =  - [R_{\#} P_2] $, so magnitude of the discrepancy will have the same distribution for sample with equal proportions of elements with distributions 
of $ R_{\#} P_1$ and $ R_{\#} P_2$. If $t^* = ( 1 - a^{-1} ) \cp$, then it can be shown that estimate $\hat \cp$ is unbiased: 
 
\begin{equation*}
\E \hat{\cp}_n = \dfrac{\sum^{T_n - \tau_n}_{t = 1} t \P\big(t = \argmin_{t' } \sccn{t'}   \big)}{1 - a^{-1}_n}
= 
\frac{ t^* +  B^+_n
     -  B^-_n  
  }
   {1 - a^{-1}_n} 
   = \frac{t^*}{1 - a^{-1}_n} = \cp  = \frac{T_n}{2},
\end{equation*}
 where
\begin{multline*}
 B^-_n = \sum^{\lfloor r(T_n - \tau_n) \rfloor}_{t = 1} t \P\Big( t^*  - t = \argmin_{t'} \sccn{t'}  \Big), \quad \\
 B^+_n = \sum^{\lfloor (1 - r)(T_n - \tau_n) \rfloor}_{t = 1}  t \P\Big(t^*  + t = \argmin_{t' } \sccn{t'}  \Big).
\end{multline*}
However, then $r \neq 1/2$ the estimate $\hat r$ is projected to  be biased towards $1/2 $ as the value 
$ B^+$ will only increase and the  value $B^-$ will only decrease  as $r$ decreases. Otherwise, increase  of $r$ 
increases the value $B^-$  too, and  decreases the value of $B^+$. 
In order to proof consistency of estimator $\hat r_n$ we introduce a family of sequences for each $s \in ((1 -a^{-1})^{-1}, 1)$
\begin{equation*}
\tilde{\Delta}^s_n =  \sccn{t'} \quad \text{having} \quad t' = \argmin_{1 \le t \le T_n - \tau_n} \left| s - \frac{t}{T_n(1 - a^{-1})} \right|
\end{equation*}
Then by the weak convergence of vector ranks for every $s$ the sequence converges to a value:
\begin{multline*}
 \tilde{\Delta}^s_n \xrightarrow[n \to \infty]{\P} \Delta^s   = 
 \Bigg\|  1_{ s < \frac{r - a^{-1} }{1 - a^{-1}}}(s)R_\# P_1 +   1_{ s > \frac{r  }{1 - a^{-1}}}(s)R_{\#} P_2 +   
 \\ +  
 1_{  \frac{r - a^{-1} }{1 - a^{-1}}\le s \le \frac{r }{1 - a^{-1}}}(s) 
 \left( \frac{ r   - s(1 - a^{-1})}{a^{-1}} R_\# P_1 +     \frac{ a^{-1} - r   + s(1 - a^{-1})}{a^{-1}} R_\# P_2 \right) \Bigg\|_{\mathcal{D}} .
\end{multline*}
Otherwise $\tilde{\Delta}^r_n \xrightarrow{\P} 0$.  We can assert by  structural uniformity of $\tilde \Delta_n$ that 
$\tilde \Delta \xrightarrow{\P} \Delta$ in Skorohod's topology.  Thus, $ \hat{r}_n = \argmin_s \Delta_n^s \xrightarrow{\P} \argmin_s  \Delta^s  = r$ providing convergence in probability.
This suggests that the bias of $\hat r_n$ can be bounded by some sequence $\beta$ with convergence $\beta_n \xrightarrow[n \to \infty]{} 0$:
\begin{equation*}
\frac{|B^+_n - B^-_n|}{T_n} \le \beta_n
\end{equation*}
Without loss of generality assume that $r < 1/2$. Then, we separate positive bias into mixing and non-mixing parts:
\begin{multline*}
\frac{|B^+_n - B^-_n|}{T_n} <  \frac{B^+_n}{T_n} < 
	\\ <
	\sum_{t = 1}^{\lfloor(1-r)\tau_n \rfloor}\frac{ t\P\Big(\sccn{t'} < \tilde \Delta^r_n \Big)}{T_n}  +  \sum_{t = \lfloor(1-r)\tau_n \rfloor + 1}^{T_n - \tau_n - t^*} \frac{ t\P\Big(\sccn{t'} < \tilde \Delta^r_n \Big) }{T_n}. 
\end{multline*}
For non-mixing part apply Markov inequality for each fixed value of $\tilde \Delta^r_n$ and some value $\zeta > 1$
\begin{equation*}
 \P\Big(\sccn{t'} < \tilde \Delta^r_n \Big) = \P\Big( \big(\sccn{t'} \big)^{-\zeta} >  \big(\tilde \Delta^r_n \big)^{- \zeta}\bigg) \le   \big( \tilde \Delta^r_n \big)^{\zeta} \E \bigg( \big(\sccn{t'} \big)^{-\zeta} \bigg). 
\end{equation*}
It is possible to use inverse as $\tilde \Delta_n > 0$ for each $n$. With this inequality  the non-mixing part can be bounded
\begin{multline*}
 \sum_{t = \lfloor(1-r)\tau_n \rfloor + 1}^{T_n - \tau_n - t^*} \frac{ t\P\Big(\sccn{t'} < \tilde \Delta^r_n \Big) }{T_n} 
 \le T_n \E \bigg(  \big( \tilde \Delta^r_n \big)^{\zeta} \bigg) \E \bigg( \big( \tilde \Delta^1_n \big)^{-\zeta} \bigg) 
 \le \\	
 \le C_r T_n \E \bigg(  \big( \tilde \Delta^r_n \big)^{\zeta} \bigg).
\end{multline*}
As $\tilde \Delta^r_n \to 0$ with the rate $O(\tau_n^{-1 + \varepsilon})$ it is  possible to specify a constant $\zeta$ in such a way that 
$T_n \E (  ( \tilde \Delta^r_n )^{\zeta} ) \downarrow 0$ and the expectation  $\E ( ( \tilde \Delta^1_n )^{-\zeta} )$ approaches some constant value by weak convergence and therefore can be bounded. 
Thus, the non-mixing bias approaches zero as $n$ goes to infinity. Note, that all moments of discrepancy exists as it is bounded on a positive interval for each $n$.
This method can be extended in order to show that the estimate is asymptotically unbiased  under an assumption of change point slackness,
\begin{definition}
Sequence of diphoragrams $\sccn{t}$ has \textbf{slackness property} iff for some constant $\gamma \in (0,a^{-1})$  and  for all $n$ large enough there are  time points $t^{**}_n > t^*_n$ such that :
\begin{equation*}
 \frac{t^{**}_n - t^*_n}{T_n} \ge \gamma, \quad  \sum_{t = 1}^{t^{**}_n} t\P
 \Big( t = \argmin_{t'} \sccn{t'}\Big) - B^-_n \le 0.
\end{equation*}
Then, it is possible to construct a similar bound
$$
\frac{| B^+_n - B^-_n|}{T_n} \le   T_n \E \Bigg(  \big( \tilde \Delta^r_n \big)^{\lambda}  \E \bigg( \big( \tilde \Delta^{\gamma + r}_n \big)^{-\lambda} \Big| \tilde \Delta^r_n  \bigg) \Bigg),
$$
which converges to zero at infinity.
\end{definition}

\subparagraph{alternative methods}
One of the negative properties of the method described in the previous chapter is the requirement of specification of bandwidth $\tau$, which prevents change point detection in the proximity of  the limit points $1$  and $T$. 
However, as space of nonuniformities  $\mathcal{D}$  is a metric space, it is possible to determine change point by maximizing distance between two empirical measures 
$ \mathrm{dist}(\hat \cp) = \| \hat \mu^+_{\hat \cp} - \hat \mu^-_{\hat \cp} \|^2_\mathcal{D}$, where 
\begin{equation}
\label{measures}
\hat \mu^-_{\hat \cp}  = \frac{1}{\widehat{\cp}} \sum^{\hat \cp}_{t = 1} \delta_{Y_t}, 
\quad
 \hat \mu^+_{\hat \cp}  = \frac{1}{T - \widehat{ \cp}} \sum^{T}_{t = \hat \cp + 1} \delta_{Y_t} .
\end{equation}
Then, by definition of the norm the distance is computed as
 \begin{multline*}
     \mathrm{dist}\Big(\hat \cp \Big) =  
  \frac{1}{(\hat \cp)^2} \sum^{\hat \cp}_{n,m = 1}   \mathcal{K}(Y_n,Y_m) - \\ 
   -\frac{2}{ \hat \cp (T - \hat \cp)}\sum^{\hat \cp}_{n = 1} \sum^{T  \cp}_{m = \hat \cp + 1} \mathcal{K}(Y_n,Y_m) 
   +  \frac{1}{( T - \hat \cp)^2} \sum^{ T }_{n,m = \hat \cp + 1}   \mathcal{K}(Y_n,Y_m),
 \end{multline*}
leading to a change point estimate $\hat \cp = \argmax_{\tilde{\cp}} \mathrm{dist}(\tilde{\cp})$. As empirical measures of $\eqref{measures}$ will converge to some points of interval $[R_\# P_1, R_\# P_2]$ as $T_n$ goes to infinity, the estimate $\hat \cp_n$ converges to true value $\cp$ in probability.

Change points detection methods of this forms were explored in the work \cite{matteson}. Thus, we will not explore it in further depth. The important properties of this statistic is that it still can be computed in $O(T^2)$ time and that $\mathrm{dist}(\hat \cp)$ is a U-statistic.

 Geometric idea of \eqref{geometric_idea} suggests that the value
 \begin{equation*}
  \varsigma\Big(\hat \cp \Big) =  \frac{\hat \cp}{T}   - \frac{\| \hat \mu^{+}_{\hat \cp}  \|_{\mathcal{D}}}{\|\hat \mu^{-}_{\hat \cp}\|_{\mathcal{D}} + \| \hat \mu^{+}_{\hat \cp}\|_\mathcal{D}}   
 \end{equation*}
 approaches zero as data size grows to infinity  iff $\hat r$ approaches the true ratio  $r$. Thus, if change point exists, then change point can be estimated as $\hat \cp = \argmin_{\tilde \cp} |\varsigma(\tilde \cp)| $.
 Or alternatively compute $\hat r^m$ by applying iteration
 $$
  \hat r^m =   \frac{\| \hat \mu^{+}_{r^{m -1}}  \|_{\mathcal{D}}}{\|\hat \mu^{-}_{r^{m-1}}\|_{\mathcal{D}} + \| \hat \mu^{+}_{r^{m - 1}}\|_\mathcal{D}}, 
 $$
 where $\mu^{+}_{r^{m - 1}}$ and $\mu^{-}_{r^{m - 1}}$ are empirical measures corresponding to the estimates obtained at previous iterations. The initial value $\hat r^0$ can be selected to be equal to $1/2$.
\paragraph{Multiple Change Points }
In this chapter the situation of $K$ possible consecutive  change points $\cp = (\cp^1,\ldots,\cp^K)$ in the data is considered. Now, $r$ denotes a list of positive values
   $$
    r =  (r^1, \ldots, r^K) = \left( \frac{\cp^1}{T}, \frac{\cp^2 - \cp^1}{T},\ldots,\frac{\cp^K - \sum^{K-1}_{k = 1} \cp^k}{T} \right), 
   $$
 which can be understood as the first $K$ coefficients in the convex combination  of $K + 1$ probability distributions $P_1, \ldots, P_{K + 1}$. That is, define
$$
\mu_r = \sum^K_{k = 1} r^k P_k  + \left( 1 -  \sum^K_{k = 1} r^k\right)P_{K + 1}.
$$
Furthermore, estimates $\hat \cp$ and $\hat r$ are treated as lists of corresponding structure. 

By definition of vector rank function $R_\# \mu_r = \Uni$.  However, in order to apply methods similar to ones developed in the previous chapter we need one more property of the model.  
\begin{definition}
$K$ probability measures $P_1, \ldots P_{K}$ are said to be \textbf{convexly independent} if for all lists of $K$ coefficient $\lambda \in \Reals^K_+$, such that $\sum^K_{k =1} \lambda_k = 1$, for each $i$ equality 
$
 P_i = \sum^K_{k = 1} \lambda_k P_k,
$
 holds only if $\lambda_j = \delta_{i,j}$ for each $j$, where $\delta_{i,j}$ is the Kronecker delta.
\end{definition}
Assume that the true value of $K \in \mathfrak{K}$. Then, by construction $ \widehat{K}^{\mathrm{SMA}} \le K$ with probability $\alpha$.

 It can be postulated, that $0$ of $\mathcal{D}$  lies in the convex hull of $R_\# P_i$. That is  
$
 0 \in \mathrm{conv} \{  R_\# P_k  \}^{K + 1}_{k = 1}. 
$
It suggests that there are projections of $0$ to the edges of the convex polytope $\pi_{k}0 \in [R_\# P_{k}, R_\# P_{k + 1}]$ which minimizes $\|  \pi_k 0   \|_{\mathcal{D}}$. Hence, if $\tau$ is taken to be small enough then local minimas of $\widehat{\Delta}^{T,\tau}$ will happen in the proximity of the change points with high probability. 

The problem with this method is that proportions of points from different sides of  a change point at the local minimise
are given by the relation
\begin{multline*}
\pi_k 0 =  \frac{\| R_\# P_{k + 1} \|^2_\mathcal{D} - \langle R_\# P_k,R_\# P_{k + 1} \rangle_{\mathcal{D}} }{d^2_{\mathcal{D}}(R_\# P_k,R_\# P_{k + 1})}R_\# P_k + 
	 \\ + 
	\frac{\| R_\# P_{k} \|^2_\mathcal{D} 
	- \langle R_\# P_k,R_\# P_{k + 1} \rangle_{\mathcal{D}} }{d^2_{\mathcal{D}}(R_\# P_k,R_\# P_{k + 1})}R_\# P_{k + 1}  ,
\end{multline*}
which can not be recovered from the diphoragram. For this reason we propose an iterative procedure. Let  $
\hat{ \mu }_{\hat{\cp}^k,\hat{\cp}^{k + 1}}^n $ denote empirical measures of points from the interval bounded by change point estimates $ [\hat{\cp}^k]^n$ and $ [\hat{\cp}^{k + 1}]^n$ .
\begin{enumerate}
\item start with $\mathcal{T}_1 = \mathcal{T} $.
\item for each $k \le K$ select $t^*_k = \argmin_{t \in \mathcal{T}_k} \tilde \Delta^{T, \tau}_{t }$ and update $T_{k + 1} = \{ t \in \mathcal{T}_k : | t - t^*_k | > \tau  \}$.
\item make initial change point estimation with blind adjustment $\Big[\hat \cp^k  \Big]_1= t^*_k + \tau/2$.
\item  readjust change points for a fixed number of iterations $N$  with nth plus one readjustment being 
 \begin{equation*}
   \Big[\hat \cp^k  \Big]_{n + 1} =  t^*_k + \Big[\hat \lambda^k_1 \Big]_n \tau,
 \end{equation*}
 where
 \begin{equation*}
   \Big[\hat \lambda^k_1 \Big]_n =  \frac{\Big\| \mu^n_{\hat \cp_{k }, \hat \cp_{k + 1} } \Big\|^2_\mathcal{D} - \Big\langle \mu^n_{\hat \cp_{k -1 }, \hat \cp_{k }},\mu^n_{\hat \cp_{k }, \hat \cp_{k + 1} } \Big\rangle_{\mathcal{D}} }{d^2_{\mathcal{D}}\Big(\mu^n_{\hat \cp_{k - 1 }, \hat \cp_{k}},\mu^n_{\hat \cp_{k }, \hat \cp_{k + 1}}\Big)}
 \end{equation*}
 with surrogate change points being $\Big[ \hat \cp^{0} \Big]_{n-1} = 0$ and  $\Big[ \hat \cp^{K + 1} \Big]_{n - 1} = T$ .
\end{enumerate}
In order to approach problem of model misspecification we apply smallest accepted model (SMA) methodology. 
For a collection of $\widehat{K}$ we acquire minimizing time points $ t^* $. Then, test for change points in the intervals bounded by $\hat \cp_i$ and $\hat \cp_{i + 1}$ with surrogate values as above by
computing a p-value approximations.

Thus, the model can be estimated while new local minimizers are being recovered and the process can be stopped as minimal accepted value of $\hat K$ has been achieved. 
\paragraph{Computational Results}
In order to conduct computational experiments the  methods discussed in the previous chapter were implemented in Python programming language with use of numpy and scipy libraries. 
We use  \texttt{sobol\_seq} package in order to  generate Sobol sequences.
\subparagraph{Simulations with Zero Change Points}
 \begin{figure}[H]
 \begin{center}
 \includegraphics[scale= 0.3]{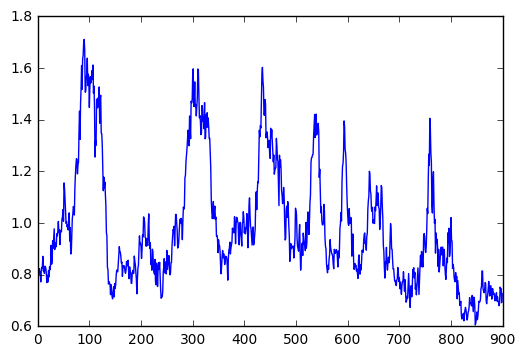} 
 \caption{Example of a control chart for 1000 observations of 5-dimensional data with standard normal distribution without change points computed by star kernel with $\tau = 100$. Abscissa represents values of shifted time $t$ in a range from 1 to 900 and ordinate stands for value of $\hat \Delta_t$.  Successfully, no change points were detected. Note, that whole process $\hat \Delta_t$ behaves as a martingale with 'small' expected value.}
 \end{center}
 \end{figure}
For simulations with zero change points we are interested in measuring statistical significance or confidence of the test statistic $T$ which can be understood as 
$$
 \mathrm{significance}(T) = \P( T \, \textrm{rejects} \, H_0 | H_0 \, \textrm{is true}), \quad
 \mathrm{confidence}(T) = \P(T \, \textrm{accepts} \, H_0 | H_0 \, \textrm{is true}).
$$
In simulations with zero change points $H_0$ obviously is true. So, for a run of $n$ simulations we estimate confidence as 
$$
\widehat{\mathrm{conf}}(T,n) = \frac{\#\{\textrm{simulations with no change points detected}  \}}{n}.
$$
We run simulations without change points and vary certain fixed parameter while measuring confidence and inverse $p$-value for each value of parameter. The only nontrivial results were obtained for change in data dimension $d$. 
 \begin{figure}[H]
 \begin{center}
 \includegraphics[scale= 0.3]{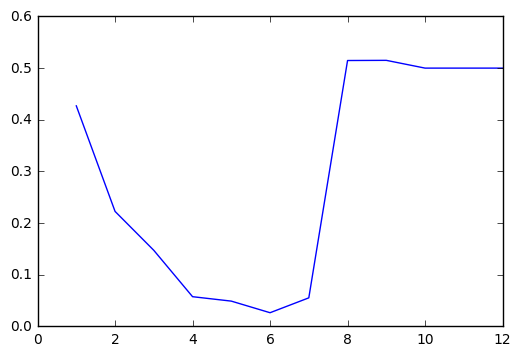} 
 \caption{Change of mean inverse $p$-value with the growth of dimensionality. We ran 20 simulations for each value $d$ of dimension in range from 1 to 12. 
Each simulation produced sample 200 standard normal variables and no change points. 
Then hypothesis were tested with significance parameter $\alpha = 0.1$ and  diphoragrams built with $\tau = 30$. 
Observe that  the inverse $p$-value decreases from $1$ to $6$ which can be understood as increase in statistical confidence in the absence of change points. 
Then it goes to the value of $0.5$ which can be interpreted as complete statistical indetermination due to complete lack of information caused by high data dimension relative to the sliding window bandwidth. }
 \end{center}
 \end{figure}
Additional experiments were conducted with growing variance and changes in covariance structure of the observations, however no dependencies were identified. This could be due to stabilizing effect of vector rank functions.

\paragraph{Simulations with One Change Point}
 \begin{figure}[H]
 \begin{center}
 \includegraphics[scale= 0.3]{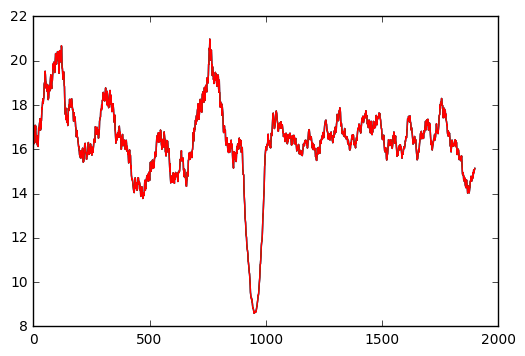} 
 \caption{Example of a diphoragram for 2000 observations of 5-dimensional data with standard normal distribution from one side of change point at $\cp =1000$ and normal distribution with mean of value 5 computed by star kernel with $\tau = 100$. 
Abscissa represents values of shifted time $t$ in a range from 1 to 1900 and ordinate stands for value of $\hat \Delta_t$.  
Successfully, change point was estimated at $\hat \cp = 1000$ with $t^* = 950$. 
Note, that the whole process $\hat \Delta_t$ behaves as a martingale with 'big' expected value in the distance of change point and goes for a 'swoop' in proximity of $\cp$.}
 \end{center}
 \end{figure}
  While running a simulations with one change point we are naturally interested in the estimation of statistical power of the test $T$ which can be understood as
  $$
   \mathrm{power}(T) = \P( T \, \textrm{rejects} \, H_0 | H_0 \, \textrm{is false}), 
  $$
and can be estimated for $n$ simulations as 
$$
 \widehat{\mathrm{pow}}(T,n) =  \frac{\#\{\textrm{simulations with change points detected}  \}}{n}.
$$
We investigate dependence of $\widehat{\mathrm{pow}}(T,n)$ on differences between distributions from opposite sides of the change point.
 \begin{figure}[H]
 \begin{center}
 \includegraphics[scale= 0.3]{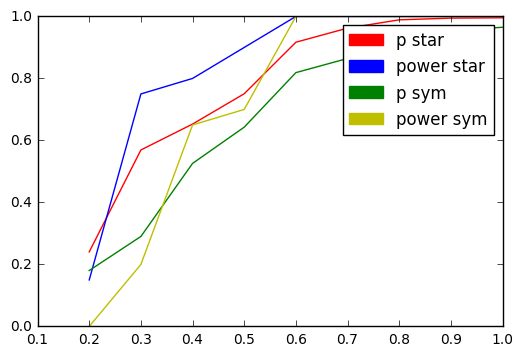} 
 \caption{Change of inverse p-value and statistical power with the growth of difference in the mean. 
	 We ran 20 simulations for each values of the difference in the range from 0.2 to 1 with the step equal to 0.1. 
	 Each simulation contained a sample of 200 
	 normally distributed variables with $d = 3$ and $\cp = 100$ with corresponding difference 
	 in the mean of distributions from both sides of the change point. The hypothesis was tested with $\alpha = 0.1$ and bandwidth $\tau = 30.$ 
	 The graph shows that the star kernel outperforms symmetric kernel at  all values of the difference. }
 \end{center}
 \end{figure}
 \begin{figure}[H]
 \begin{center}
 \includegraphics[scale= 0.3]{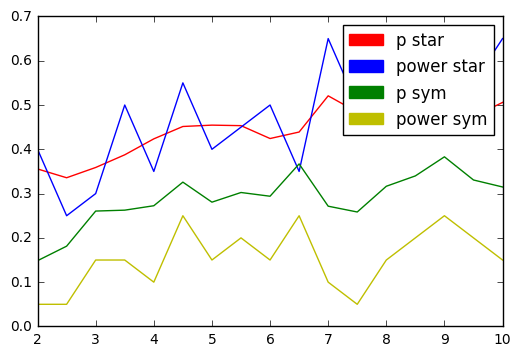} 
 \caption{Change of inverse p-value  ans statistical power with the growth of difference in the variance. We ran 20 simulations for each values of the difference in the range from 2 to 10 with the step equal to 0.5. 
	 Each simulation contained a sample of 200 normally distributed variables with $d = 5$ and $\cp = 100$ with corresponding difference in the variance of distributions from both sides of the change point. 
	 The hypothesis was tested with $\alpha = 0.1$ and bandwidth $\tau = 30.$ The graph shows that the star kernel outperforms symmetric kernel at all  values of the difference. }
 \end{center}
 \end{figure}
 
 The figures shows that star discrepancy outperforms symmetric discrepancy  in detecting change both in expectation and in variance. However empirical results in \cite{liang} indicates that in some situations symmetric discrepancy may turn out tob be a better tool.

In the situation of existing change point not only power of the test is of interest but also a precision of change point estimations. 
As it was shown in previous chapter bias of a change point estimate increases as true ratio $r$ of  distributions  in the sample diverges from the value of $1/2$. 
We provide a computational illustrations:

 \begin{figure}[H]
 \begin{center}
 \includegraphics[scale= 0.3]{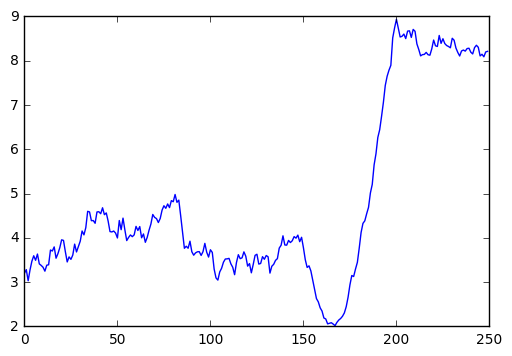} 
 \caption{ Discrepancy diphoragram for 300 normal i.i.d  observations with $d = 3$ with change in expectation at $\cp = 200$. 
	 Abscissa represents values of shifted time $t$ in a range from 1 to 250 and ordinate stands for value of $\hat \Delta_t$ constructed with $\tau = 50$. 
	 Change point was detected at location $\hat \cp = 197$, which can be interpreted as a weak drift towards  midpoint. 
	 Note, that the relation of mean values in martingale parts of $\hat \Delta_t$ at different sides of the change point corresponds 
	 to the nonuniformity space theory presented in the previous chapter.  }
 \end{center}
 \end{figure}
  \begin{figure}[H]
 \begin{center}
 \includegraphics[scale= 0.3]{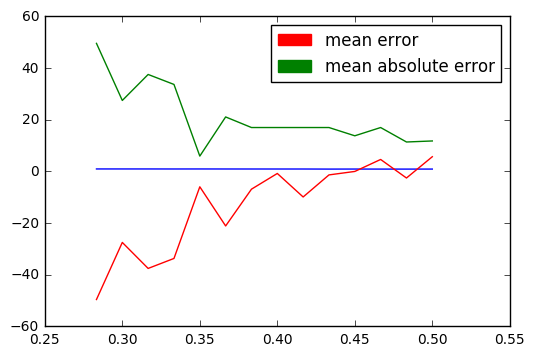} 
 \caption{ Change of error $\cp - \hat \cp$ and  absolute  error $|\cp - \hat \cp |$  with shift of value of $r$. We ran 20 simulations for each values of the ration in the range from $1/2$ to $13/60$ with the step equal to $1/60$. 
	 Each simulation contained a sample of $r*300$ normally distributed variables with zero expectation at one side of change point and $(1 - r)*300$  variables uniformly distributed in $[-1,1]$ square  with $d = 2$ and $\cp = r*300$ . 
	 Change points were estimated with $\tau = 50$. 
	 The graph shows that error indeed grows in the distance of the midpoint and that estimate has strong drift towards $1/2$ for such ratios.   }
 \end{center}
 \end{figure}
\subparagraph{Simulations with Multiple Change Points}
For case of multiple change point we provide only example with diphoragrams:
  \begin{figure}[H]
 \begin{center}
 \includegraphics[scale = 0.3]{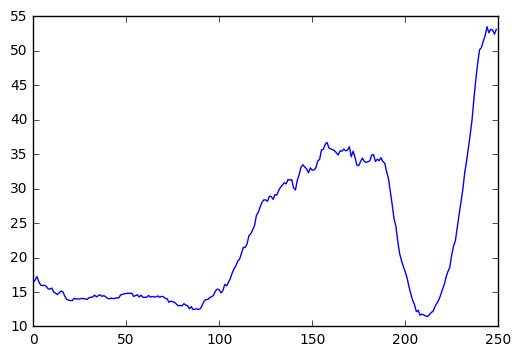} 
 \caption{ Discrepancy diphoragrams for 300 observations with $d = 3$.  
	 Sample has distribution model $P_1 = \Normal((3,3,3),I), P_2 = \mathcal{U}[10,20]^3$ and $P_3 =  \Normal(-(3,3,3),I)$  with  $\cp_1 = 120$ and $\cp_2 = 240$. 
	 Abscissa represents values of shifted time $t$ in a range from 1 to 250 and ordinate stands for value of $\hat \Delta_t$ constructed with $\tau = 50$. 
	 Change point was detected at locations $\hat \cp_1 = 111$ and $\hat \cp_2 = 237$. However, after application of iterative correction improved estimate $\hat \cp_1' = 119 $ and $\hat \cp_2' = 241$ were acquired.}
 \end{center}
 \end{figure}
\paragraph{Examples with financial data:}
In order to provide examples, which are not artificially constructed, financial data similar to one in \cite{finance}. 
This data was acquired from \href{http://mba.tuck.dartmouth.edu/pages/faculty/ken.french/data_library.html}{Data library}  of  Keneth R. French. It contains mean monthly returns from five portfolios each composed of one of  five industrial sectors in the USA, which are:
\begin{enumerate}[label=(\Alph*)]
	\item Finance,
	\item Manufacturing,
	\item Retail, wholesale  and some services,
	\item Utilities,
	\item Other.
\end{enumerate}
This provides data dimension of $d = 5$ and total number of  observations of $T = 1091$ as they were recorded monthly from July of the year 1925 to the may of the year 2017.   
\begin{figure}[H]
\begin{center}
(A) \includegraphics[scale = 0.22]{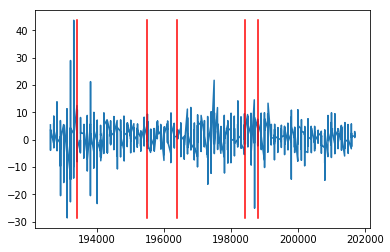} 
(B) \includegraphics[scale = 0.22]{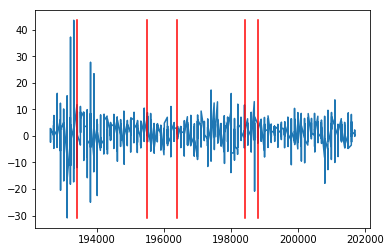} 
(C) \includegraphics[scale = 0.22]{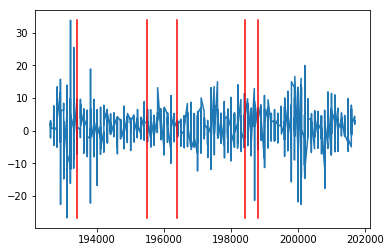}
\\
(D) \includegraphics[scale = 0.22]{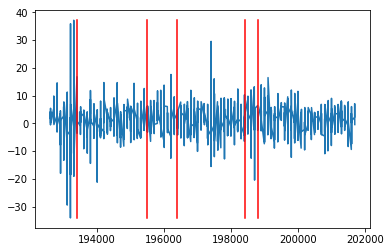}
(E) \includegraphics[scale = 0.22]{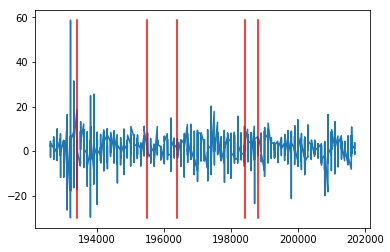}  
\caption{
	We run SMA-based change point estimation with star kernel,bandwidth equal to 50 and the false alarm rate equal to 0.2. The procedure  provides 5 change points at moment
	March of 1934, April of 1955, April of 1964, December of 1984, September of 1988.   
}
\end{center}
\end{figure}
In the original paper \cite{finance} the parametric Bayesian inference was used too estimate nine and seven change points, and the total number of change point was only guessed and not inferred. 
Our results differ significantly from this original results, however different time range was used.      
\begin{figure}[H]
\begin{center}
	
	\begin{subfigure}[t]{0.03\textwidth} (A) \end{subfigure}
	\begin{subfigure}[t]{0.27\textwidth}
	\includegraphics[scale = 0.22]{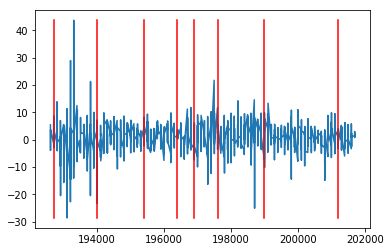} 
	\end{subfigure}
	\begin{subfigure}[t]{0.03\textwidth}(B)\end{subfigure}
	\begin{subfigure}[t]{0.27\textwidth}
	\includegraphics[scale = 0.22]{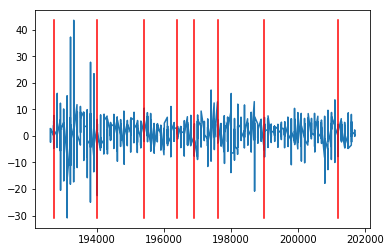} 
	\end{subfigure}
	\begin{subfigure}[t]{0.03\textwidth}(C) \end{subfigure} 
	\begin{subfigure}[t]{0.27\textwidth}
	\includegraphics[scale = 0.22]{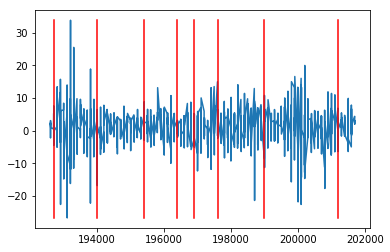}
	\end{subfigure}
\\
	(D) \includegraphics[scale = 0.22]{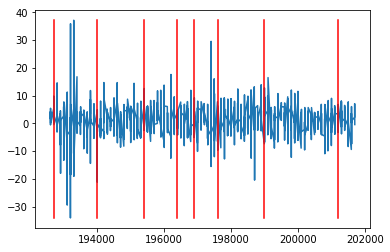}
	(E) \includegraphics[scale = 0.22]{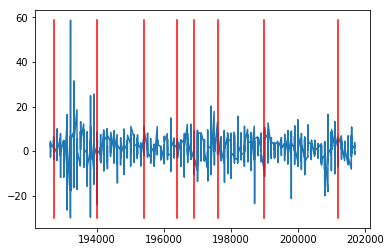}  
\caption{
	By running SMA-procedure with symmetric kernel, bandwidth equal to 40 and the false alarm rate of 0.2, we get less conservative estimate. 
	Change points are February 1927, October 1940, July 1954, April 1964, January 1969, May 1976, May 1990, July 2012. 
}
\end{center}
\end{figure}
It can be projected that all above change points can be attributed to the important events in the economic history. For example, change point at July 1954 can be related to the end of recession of 1953, 
which itself can be explained by the change of interrelations  of the industrial sectors leading to the change in the structure of the observed distribution. 
Hence, we can describe performance of the SMA-procedure as satisfactory.

\paragraph{Conclusion and Discussion}

As result of the work a collection of methods of change point detection methods was developed. All this methods are based on interaction of vector ranks and geometric discrepancy which is a novel result. Certain basic consistency result were proved for the method in its basic form designed for detecting a single change point. However, they also can be applied for detecting and estimating multiple change points. 

Computational results shows applicability of the method both for simple artificial change point problems an problems concerning actual data from applications. 
It is strictly indicated by resulting experience that the method is much more potent in situations then the distributions are not concentric. 
Empirical  evidences of our theoretical findings were also observed. 
These theoretical results include expression of relation on different sides of change point through inner product in the Hilbert space $\mathcal{D}$. 
Another Theoretical result concerns dependence of estimate's bias on ratio in which change point separates sample. 

Another positive results is the discovery of $\mathcal{D}$, which can be used for establishing alternative iterative  procedures defined by relations in this Hilbert space. Furthermore, this Hilbert space structure on empirical measures can be used for proving more theoretical results in the future.  

Better proofs which cover ergodic processes and provide exact rates of convergence a still need to be worked out for the methods. 
Furthermore concentration results for change point estimates might render this algorithms interesting for practical application. 
However, they are absent in the current work.

Finally, online version of method can be implemented. The only requirement for this algorithm is fast online computation of the optimal assignment problem. 
It can be projected that such algorithm can be  derived from the Sinkhorn distance regularised algorithm which were designed by Cuturi \cite{Kuturi}. 
\printbibliography
\end{document}